\documentclass[twocolumn,english,superscriptaddress]{revtex4-1}

\usepackage[T1]{fontenc}
\usepackage[latin9]{inputenc}
\usepackage{graphicx}
\usepackage{amsmath}
\usepackage[colorlinks = true,
	linkcolor = blue,
	urlcolor  = blue,
	citecolor = blue,
	anchorcolor = blue]{hyperref}
\usepackage[usenames, dvipsnames]{color}

\makeatletter
\usepackage{babel}
\usepackage{float}
\usepackage[normalem]{ulem}

\makeatother

\begin{document}

%%%% This disclaimer is mandatory for all ORNL preprints
%\setcounter{page}{0}
%Notice: This manuscript has been authored by UT-Batelle, LLC, under contract DE-AC05-00OR22725 with the US Department of Energy (DOE). The US government retains and the publisher, by accepting the article for publication, acknowledges that the US government retains a nonexclusive, paid-up, irrevocable, worldwide license to publish or reproduce the published form of this manuscript, or allow others to do so, for US government purposes. DOE will provide public access to these results of federally sponsored research in accordance with the DOE Public Access Plan (http://energy.gov/downloads/doe-public-access-plan).
%
%\newpage
%
%\quad
%
%\newpage

\title{
  Quantum wake dynamics in Heisenberg antiferromagnetic chains
  %% Real-time dynamics and Floquet states in quantum magnetic chains measured by neutron scattering
}

\author{A. Scheie}
\affiliation{Neutron Scattering Division, Oak Ridge National Laboratory, Oak Ridge, Tennessee 37831, USA}

\author{P. Laurell}
%\affiliation{Center for Nanophase Materials Sciences, Oak Ridge National Laboratory, Oak Ridge, TN 37831, USA}
\affiliation{Computational Sciences and Engineering Division, Oak Ridge National Laboratory, Oak Ridge, TN 37831, USA}
\affiliation{Department of Physics and Astronomy, University of Tennessee, Knoxville, TN 37996, USA.}

% \author{A. M. Samarakoon}
% \affiliation{Neutron Scattering Division, Oak Ridge National Laboratory, Oak Ridge, Tennessee 37831, USA}

\author{B. Lake}
\affiliation{Helmholtz-Zentrum Berlin f{\"u}r Materialien und Energie GmbH, Hahn-Meitner Platz 1, D-14109 Berlin, Germany}
\affiliation{Institut f{\"u}r Festk{\"o}rperphysik, Technische Universit\"at Berlin, Hardenbergstra{\ss}e 36, D-10623 Berlin, Germany}

\author{S. E. Nagler}
\affiliation{Neutron Scattering Division, Oak Ridge National Laboratory, Oak Ridge, Tennessee 37831, USA}
\affiliation{Quantum Science Center, Oak Ridge National Laboratory, Tennessee 37831, USA}

\author{M. B. Stone}
\affiliation{Neutron Scattering Division, Oak Ridge National Laboratory, Oak Ridge, Tennessee 37831, USA}

\author{J-S Caux}
\affiliation{Institute of Physics and Institute  for  Theoretical  Physics,  University  of  Amsterdam, PO  Box  94485,  1090  GL  Amsterdam,  The  Netherlands}

\author{D. A. Tennant}
\affiliation{Neutron Scattering Division, Oak Ridge National Laboratory, Oak Ridge, Tennessee 37831, USA}
\affiliation{Quantum Science Center, Oak Ridge National Laboratory, Tennessee 37831, USA}
\affiliation{Shull Wollan Center - A Joint Institute for Neutron Sciences, Oak Ridge National Laboratory, TN 37831. USA}

\date{\today}

%\begin{abstract}
%\end{abstract}

\maketitle

%%% Abstract should have citations inside it.
\textbf{
Traditional spectroscopy, by its very nature, characterizes properties of physical systems in the momentum and frequency domains. The most interesting and potentially practically useful quantum many-body effects however emerge from the deep composition of local, short-time correlations. Here, using inelastic neutron scattering and methods of integrability, we experimentally observe and theoretically describe a local, coherent, long-lived, quasiperiodically oscillating magnetic state emerging out of the distillation of propagating excitations following a local quantum quench in a Heisenberg antiferromagnetic chain. This ``quantum wake'' displays similarities to Floquet states, discrete time crystals and nonlinear Luttinger liquids.
}\\
%\section{Introduction}

Ever since its introduction, the Heisenberg chain~\cite{1928_Heisenberg_ZP_49}
\begin{equation}
\mathcal{H}=J\sum_{i=1}^{N}{\Vec{S}_i}\cdot{\Vec{S}_{i+1}}    \label{eq:Heisenberg}
\end{equation}
has been the paradigmatic model of strongly-correlated many-body quantum physics. Its exact solution by Bethe~\cite{1931_Bethe_ZP_71} gave birth to the field of quantum integrability;
its magnetic excitations, spin-$1/2$ spinons~\cite{1981_Faddeev_PLA_85}, are the
prototypical fractionalized excitations.
%Perhaps the most pleasantly lasting aspect of the model is that it provides an accurate description of real
The model is not simply a theoretical archetype, but also effectively describes many physical quantum 
magnets such as KCuF$_3$~\cite{Tennant1993,Lake2013}, in which
the chains are formed by magnetic Cu$^{2+}$ ions hybridizing along the $c$ axis. Although KCuF$_3$ orders magnetically at $T_n = 39$~K, even below the ordering temperature its high energy spectrum retains the characteristic spinon spectrum~\cite{Lake2005_PRB} while exhibiting strong quantum entanglement~\cite{scheie2021entanglement}.

One of the best experimental tools for studying magnetic excitations is inelastic neutron scattering~\cite{Brockhouse_1957}, which
measures the energy-resolved Fourier transform of the space- and time-dependent spin-spin correlation function $G(r, t) = \langle S^\alpha_i(0) S^\alpha_{i+r}(t) \rangle$, ($\alpha = x,y,z)$~\cite{Squires}. Accordingly, scattering cross section data is typically reported in terms of reciprocal space and energy.
As pointed out by Van Hove in 1954~\cite{VanHove_1954,VanHove_1954_FM}, with enough data one can take the inverse Fourier transform and obtain the spin correlations in real space and time with atomic spatial resolution and time resolution of $\sim 10^{-14}$~s.
%This transformation, performed by Brockhouse himself for neutron scattering data of liquid lead~\cite{Brockhouse_1959},
This transformation was shortly thereafter applied to liquid Lead neutron scattering data~\cite{Brockhouse_1959}, and more recently on water using inelastic xray scattering~\cite{Iwashitae_2017}
but has not been applied to magnetic materials.

Space-time dynamics in one dimension has been the subject of extensive study in recent decades \cite{GiamarchiBOOK}, with attention mostly focusing on ballistically-propagating excitations (describable using bosonization / Luttinger liquid theory \cite{1981_Haldane_JPC_14}) forcing ``light-cone''-induced bounds on velocity of correlations and entanglement spreading \cite{lieb1972finite}.
The physics of Heisenberg chains is however much richer, containing nonlinearities whose effects can
be captured exactly using integrability, or asymptotically using nonlinear Luttinger liquid theory \cite{2012_Imambekov_RMP_84}.

In this paper, we use high-precision INS data transformed back to real, atomic-level space and time to characterize magnetic dynamics at the local level in a Heisenberg chain.
We focus on previously-overlooked features of the real-space/time magnetic Van Hove correlation function $G(r,t)$, namely the effects of long-term coherent, non-propagating excitations (beyond the reach of bosonization).
We observe a correlated time-dependent state resulting from the integrability-induced
``persistent memory'' of the Heisenberg chain. This state is reminiscent of a local many-body
Floquet state or a discrete time crystal,
in that it displays a characteristic time-repeating pattern with fixed period.
Its correlations also display a remarkable (spatial) ``period doubling'' (mirroring the time period doubling of a discrete time crystal), in that the original site-alternating N{\'e}el order of the
initial state changes to a two-site-spaced, oscillating antiferromagnetic correlation.
This state, which we call a ``quantum wake'' due to its similarity to the wake created by a moving ship, is a coherent wavepacket of ``deep'' and ``edge'' spinons stabilized and made observable via a Van Hove singularity, and recalls the quantum dynamical impurity picture of nonlinear Luttinger liquid theory.
% It is long-lived on the scale of magnetic exchange timescales, but remains ephemeral as a consequence of the inevitability of Anderson's orthogonality catastrophe~\cite{1967_Anderson_PRL_18}.

\section*{Results:}

\begin{figure*}
  \centering
  \includegraphics[width=0.77\textwidth]{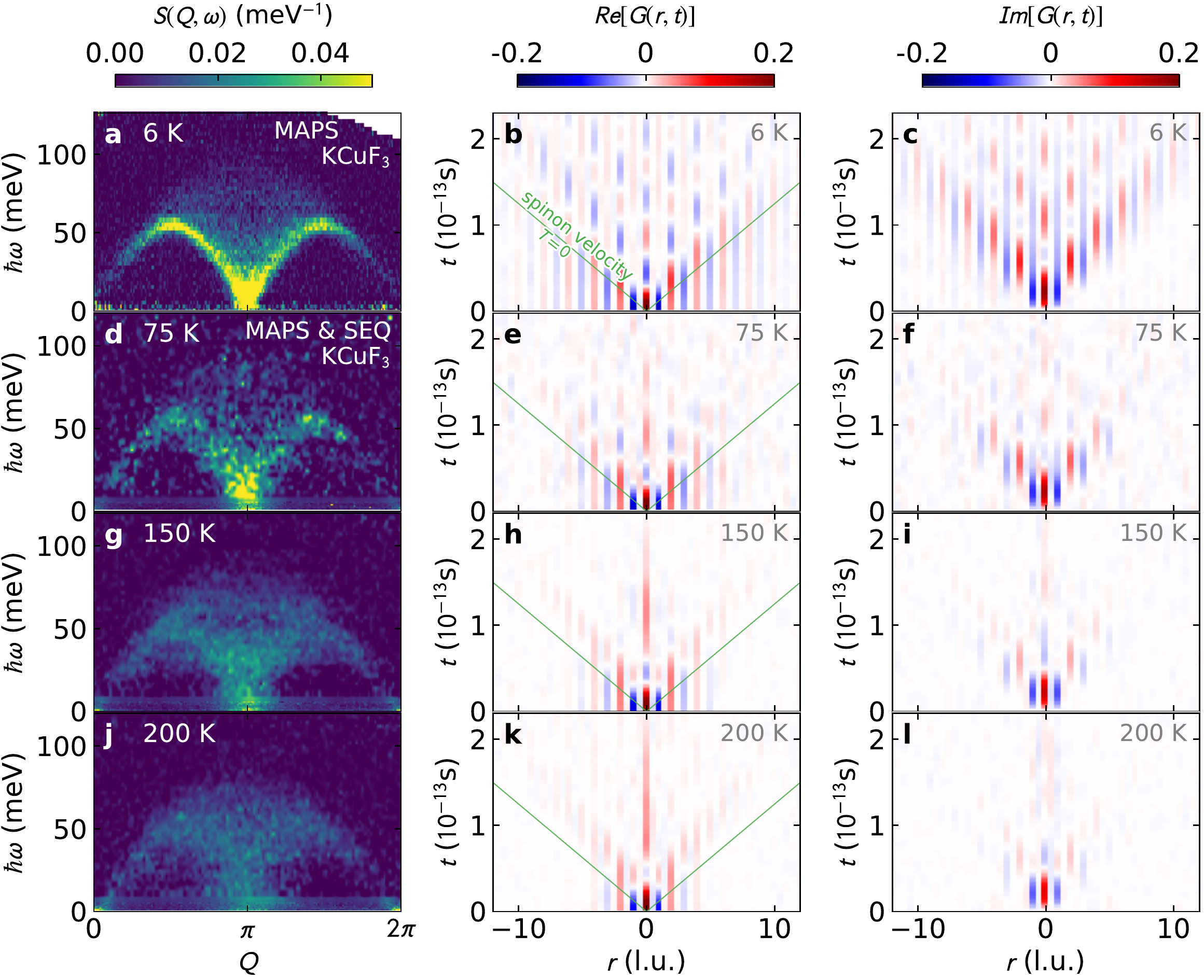}
 \caption{\textbf{Scattering and Van Hove correlations.} Finite temperature neutron    scattering data for KCuF$_3$ (left column) and their transformation to real-space correlations, with the real $G(r,t)$ (center column) and imaginary $G(r,t)$ (right column). Red indicates ferromagnetic spin correlation, blue indicates antiferromagnetic spin correlation. At low temperatures, the real $G(r,t)$ wavefront at the light cone is antiferromagnetic, and by 200 K it becomes ferromagnetic. Meanwhile, the imaginary $G(r,t)$ is restricted in space and time at higher temperatures, showing loss of quantum coherence.}
	\label{fig:hightemp}
\end{figure*}

\begin{figure*}
	\centering
	\includegraphics[width=\textwidth]{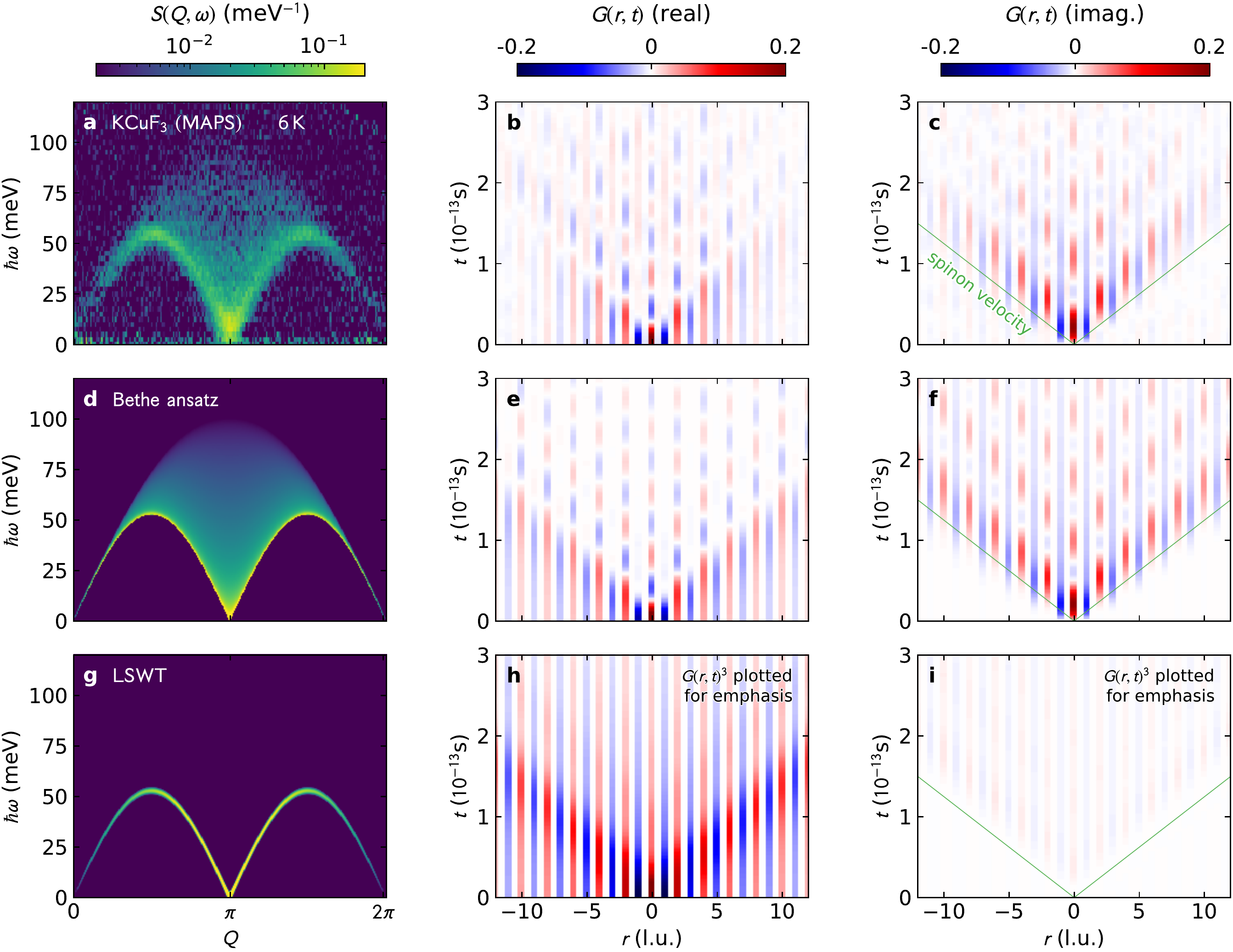}
	\caption{\textbf{Van Hove time-dependent real-space spin-spin correlation compared to theory with imaginary components. \sf a} 6~K KCuF$_3$ scattering, \textbf{\sf b} real component of $G(r,t)$, \textbf{\sf c} imaginary component of $G(r,t)$. Panels \textbf{\sf d} - \textbf{\sf f} show the same for $T=0$ Bethe ansatz, and \textbf{\sf g} - \textbf{\sf i} show the same for $T=0$ LSWT on a $S=1/2$ HAF chain (renormalized by $\pi/2$ to match the light cone velocity in the top two panels). The thin green lines on $G(r,t)$ plots show the magnon/spinon velocity.}
	\label{fig:fouriertransform_imag}
\end{figure*}

Experimental $G(r,t)$ results are obtained
using available KCuF$_3$ data from Refs.~\cite{Lake2013, scheie2020detection}
(full details are provided in the Methods section).
The result is shown in Fig.~\ref{fig:hightemp},
where ferromagnetic $G(r,t)$ correlations are shown in red and antiferromagnetic
correlations are shown in blue.
To help interpret the experimental $G(r,t)$, we also calculated $G(r,t)$ from: (i) Bethe Ansatz~\cite{Lake2013} for zero temperature, and (ii) semiclassical linear spin wave theory (LSWT).
These are shown in Fig.~\ref{fig:fouriertransform_imag}.

% \begin{figure}
%   \centering
%   \includegraphics[width=0.47\textwidth]{FourierTransform_KCuF3_HighTemp}
%   \caption{\textbf{Scattering and Van Hove correlations.} Finite temperature neutron scattering data for KCuF$_3$ (left column)
%     and their transformation to real-space correlations (right column). Red indicates ferromagnetic spin correlation, blue indicates antiferromagnetic spin correlation.
%     Only the real part of $G(r,t)$ is plotted (see Fig.~\ref{fig:fouriertransform_imag}
%     for the imaginary part). At low temperatures, the wavefront at the light cone is antiferromagnetic, and by 200 K it becomes ferromagnetic.}
% 	\label{fig:hightemp}
% \end{figure}

Real space $G(r, t)$ for spin systems can also be probed with cold atom and trapped ion experiments~\cite{Langen2013,cheneau2012light,jurcevic2014quasiparticle}, but $G(r, t)$ derived from neutron scattering has several unique advantages: (i) The systems probed by neutrons are thermodynamic, and temperature is a well-defined quantity. (ii) Neutrons explore the spin system's evolution following a local perturbation. (iii) As we show below, neutron scattering accesses the imaginary $G(r, t)$ which reveals quantum coherence and Heisenberg uncertainty.

The Fourier transform of the $S(Q, \omega)$ scattering data produces a $G(r,t)$ with
complex values, with a distinct interpretation for the real and imaginary parts. As noted by Van Hove~\cite{VanHove_1954_FM}, the imaginary part $Im[G(r,t)] = \frac{1}{2i}\langle [S_i^{\alpha}(0), S_{i+r}^{\alpha}(t)] \rangle$ (${\alpha} = x,y,z$) quantifies the imbalance between positive and negative energy scattering. By Robertson's relation \cite{Robertson1929}, a nonzero commutator between observables implies Heisenberg uncertainty; thus nonzero imaginary $G(r,t)$ indicates the presence of an uncertainty relation between $S_i^z(0)$ and $S_j^z(t)$. This mutual incompatibility is thus an indicator of quantum coherence between spins (see supplemental information). It is striking that the quantum coherence can be tracked as a function of temperature with the imaginary $G(r,t)$ in Fig.~\ref{fig:hightemp}. As temperature increases, the nonzero imaginary $G(r,t)$ shrinks to shorter and shorter times and distances, showing how the finite-temperature macroscopic world emerges from the quantum world.
On the other hand, the real part $Re[G(r,t)] = \frac{1}{2}\langle \{S_i^{\alpha}(0), S_{i+r}^{\alpha}(t)\} \rangle$
extracts classical behaviour surviving even at infinite temperature.

The real space correlations in Figs.~\ref{fig:hightemp} and~\ref{fig:fouriertransform_imag} emerge from a flipped spin at $t=0$, $r=0$.
A number of things can be observed from these $G(r,t)$ data: first, the characteristic ``light cone'' defined by the spinon velocity $v = \frac{\pi J}{2}$ where $J$ is the exchange interaction. At low temperatures, everything below the light cone is static while everything above it is dynamic.
Second, at low temperature in $G(r,t)$ there is a clear distinction between even and odd sites: the odd neighbor correlations quickly decay to zero above the light cone, whereas the even neighbor correlations persist to long times.
Third, as temperature increases the spin oscillations above the light cone shrink to shorter distances and times, until by 200~K the on-site ($r=0$) correlation oscillates only once and no neighbor-site oscillations are visible.
Fourth and finally, the wavefront above the light cone changes to ferromagnetic at high temperatures (Fig.~\ref{fig:hightemp}{\bf h}) whereas it was antiferromagnetic at low temperatures. This accompanies the nonzero imaginary $G(r,t)$ shrinking to shorter and shorter times and distances as temperature increases.

To gain a better understanding of the signal, we should identify which excitations are responsible for which part. %The famous ``viking helmet'' shape of $S(Q,\omega)$ (most clearly seen in Fig.~\ref{fig:fouriertransform_imag} panel d) shows a dominant continuum of states bound by lower $\omega_l(Q) = \frac{\pi J}{2}|\sin Q|$ and upper $\omega_l(Q) = \pi J |\sin Q/2|$ boundaries delimiting the two-spinon continuum. The lower boundary has one spinon ``sitting at the edge'' of the Fermi see while the second ``dives into'' the sea (carrying all momentum and energy), reaching the deepest point at $Q = \pi/2$; at the upper boundary, the two spinons ``dive together'', sharing momentum and energy equally.
The light cone is due to the low-energy correlations around $Q \sim \pi$ which can
be understood from traditional bosonization, the Fermi velocity being given by the group velocity of $Q \sim \pi$ spinons. % in the vicinity of the edge. 
These being the fastest-moving ballistic particles, they limit the velocity of energy, correlations, and entanglement propagation, giving the Lieb-Robinson bound~\cite{lieb1972finite,Bravyi_2006}. Such a light cone is seen in theoretical simulations~\cite{Calabrese_2007,Bonnes_2014,Collura_2015,dePaula_2017,Langer_2011,Vlijm_2016} and cold-atom experiments~\cite{cheneau2012light}, and nicely also here in KCuF$_3$.

Letting the fast-moving ballistic particles ``distill'' away leaves a ``quantum wake''
behind the wavefront, a persistent oscillating state above the light cone which is clearly seen in Fig.~\ref{fig:hightemp} panel b and Fig.~\ref{fig:fouriertransform_imag}
panels b, c, e and f. This originates from another crucial characteristic of  $S(Q,\omega)$, namely that its correlation weight is spread nontrivially within the spinon continuum. Contrasting
LSWT with Bethe Ansatz in the second and third row of Fig.~\ref{fig:fouriertransform_imag} shows stark differences in dephasing
behaviour. LSWT, being inherently coherent, has very slow dephasing and no quantum wake. For the experimental and Bethe Ansatz $G(r,t)$ however, there exist
pockets of states around $Q \sim \pi/2, \>3\pi/2$, $\omega \simeq \pi J/2$
which display a Van Hove singularity in their density of states. %; moreover, the spin correlation has a square-root singularity when approaching the lower edge $S(Q, \omega) \sim 1/\sqrt{\omega - \omega_l(k)}$~\cite{1997_Karbach_PRB_55} (neglecting logarithmic corrections), and this divergence empowers these states to give a measurable signal. The result is clearly seen in the images, with long-lived oscillations decaying as $1/\sqrt{t}$ at large times. 
Since the existence and sharpness of the lower edge are contingent on
integrability, measuring the (slowness of the) time decay of the quantum wake
is in fact a direct experimental measurement of the proximity to integrability.

\begin{figure}
	\centering
	\includegraphics[width=0.46\textwidth]{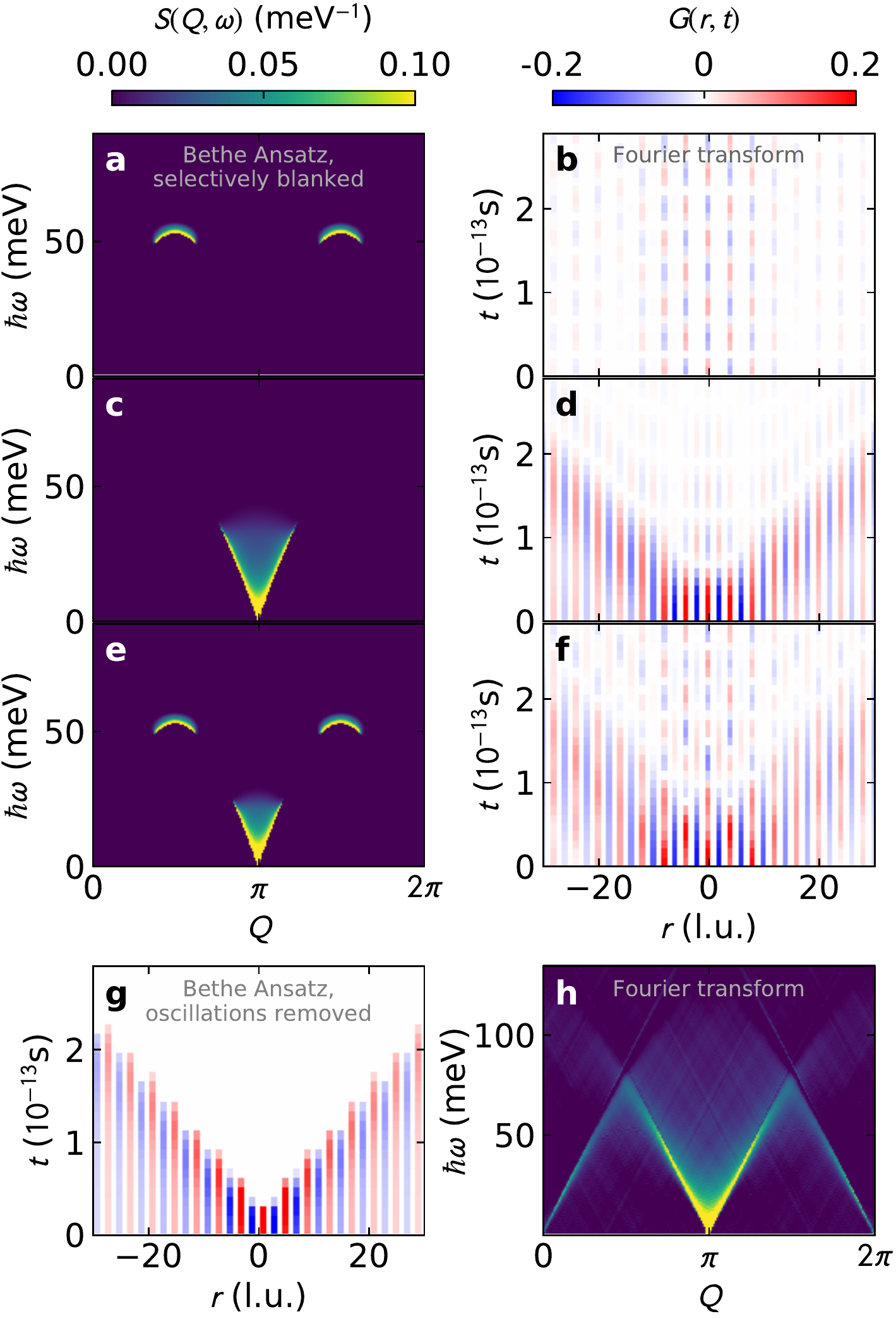}
	\caption{\textbf{Signal analysis of the Bethe Ansatz.} The right column is the Fourier Transform of the left. Panels (a), (c), and (e) show the Bethe Ansatz with everything removed but key features at $Q=\pi$ or $Q=\pi/2$. Panels (b), (d), and (e) show the resulting Fourier transform of these spectra into real space and time. This clearly shows that the oscillations above the light cone are due to the stationary $Q=\pi/2$ states, while the light cone is due to the dispersive $Q=\pi$ state. Panel (g) shows the $G(r,t)$ of the Bethe Ansatz with all correlations above the light cone set to zero. Fourier-transforming this back into $S(Q,\omega)$ in panel (h), we find a spinon spectrum with the $Q=\pi/2$ stationary states missing---confirming that these are responsible for the oscillating Floquet dynamics.}
	\label{fig:signalAnalysis}
\end{figure}

To more illustratively map the features in $G(r,t)$ with specific spinon states, we selectively remove parts of the Bethe Ansatz  $S(Q,\omega)$ spectrum, keeping only key features, and Fourier transform into  $G(r,t)$. As shown in Fig.~\ref{fig:signalAnalysis}(a)-(b), the oscillations above the light cone come from the $Q=\pi/2$ Van Hove singularities at the top of the spinon dispersion where the spinons have zero group velocity. Meanwhile, Fig.~\ref{fig:signalAnalysis}(c)-(d) shows the light cone emerges from the strongly dispersing low-energy $Q=\pi$ states. Combining these two states in Fig.~\ref{fig:signalAnalysis}(e)-(f) gives a rough reproduction of the actual $G(r,t)$, indicating that the $Q=\pi/2$ and $Q=\pi$ spinon states are what give the Heisenberg chain quantum wake its distinctive properties.
Bolstering this conclusion is the analysis shown in Fig.~\ref{fig:signalAnalysis}(g)-(h) where we remove the oscillations above the light cone from $G(r,t)$, and transform back into  $S(Q,\omega)$. In this case, we see the familiar spinon spectrum, but with the stationary $Q=\pi/2$ states missing---showing that the flat singularity at the top of the spinon dispersion is responsible for the long-lived oscillating spin correlations.

\paragraph{Quantum scrambling:}
Perhaps the most striking feature of the KCuF$_3$ quantum wake is the total loss of N{\'e}el correlations above the light cone.
Below the light cone, the system shows static $Q=\pi$ antiferromagnetism. Above the light cone, the system shows dynamic period-doubled $Q=\pi/2$  antiferromagnetism, with hardly a trace of the original state.
In stark contrast to this, equal-time real space correlators $\langle S^\alpha_i(t) S^\alpha_j(t) \rangle$ (as opposed to dynamical correlator $G(r,t)$ which measures $\langle S^\alpha_i(0) S^\alpha_j(t) \rangle$) computed from Bethe ansatz show rapid reemergence of $Q=\pi$ antiferromagnetism above the light cone, where nearest neighbor $\langle S^\alpha_0(t) S^\alpha_1(t) \rangle \rightarrow \frac{1}{12} - \frac{\ln 2}{3} \simeq -0.1477...$ ~\cite{1938_Hulthen_AMAF_26A} as $t \rightarrow \infty$.
% However, this is not at all seen in Fig.~\ref{fig:fouriertransform_imag}: $G(r=1,t) \rightarrow 0$ almost immediately, never returning to the $Q=\pi$ state.
At first glance, these results are contradictory; but the difference between $\langle S^\alpha_0(0) S^\alpha_1(t) \rangle$ and $\langle S^\alpha_0(t) S^\alpha_1(t) \rangle$ indicates the new AFM correlations form in a basis orthogonal to the original basis. In other words, the $t \rightarrow \infty$ state has zero correlations with the $t=0$ state, in accord with Anderson's orthogonality catastrophe \cite{1967_Anderson_PRL_18}.

This process can be more precisely described as quantum scrambling: the delocalization of quantum information over time~\cite{Luitz_2017,Swingle2018}. Typically such physics is studied via out of time order correlators (OTOC---see Supplemental Materials section for details). $G(r,t)$ provides an alternative and more experimentally accessible way to study quantum scrambling, quench dynamics, and quantum thermalization in physical systems.

\begin{figure}
	\centering
	\includegraphics[width=0.46\textwidth]{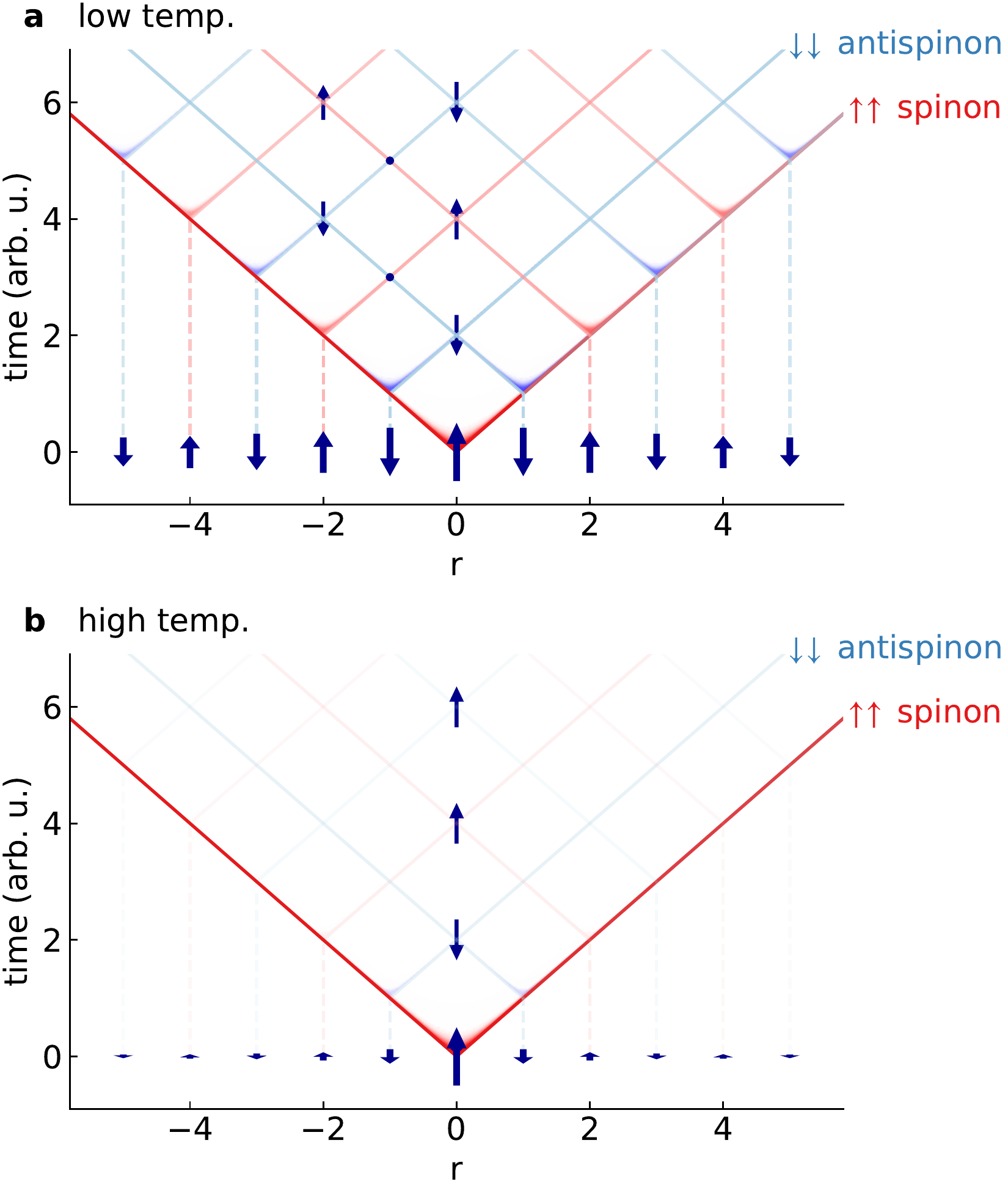}
	\caption{\textbf{Schematic description of the AFM Van Hove correlations.} At low temperatures, (a) a central spinon light cone emanates from $r=0$, $t=0$. As it reaches each neighboring site, it excites a pair of spinons which creates its own light cone. Odd neighbor sites have opposite spin from $r=0$ at $t=0$, and thus they create antispinon pairs. For even $r$, these spinon light cones create constructive interference and continue to flip spins up and down. For odd $r$, the spinons and antispinons destructively interfere, such that the correlations quickly go to zero.
		At high temperatures (b), the spin correlations are much weaker, such that the spinon and antispinon light cones emanating from $|r|>0$ are weakly coherent with $r=0$ and thus their influence is suppressed, leading to oscillations restricted in both space in time as seen in Fig.~\ref{fig:hightemp}.}
	\label{fig:schematic}
\end{figure}

\paragraph{Heuristic understanding of \texorpdfstring{$G(r,t)$}{G(r,t)}:}
The $\pi/2$ oscillations inside the quantum wake can be understood heuristically as particle-antiparticle annihilation.
In an antiferromagnetic chain, a down spin flipped up creates two spinons, while an up spin flipped down creates two antispinons. These quasiparticles interfere as schematically shown in Fig.~\ref{fig:schematic}. Spinons from even neighbor sites interfere constructively and produce a full spin flip, while antispinons from odd neighbor sites interfere destructively and annihilate. Thus $G(r,t)$ oscillates on even sites and $Re[G(r,t)]=0$ on odd sites.
%Schematically, the oscillation pattern can be understood as shown in Fig.~\ref{fig:schematic}. As a single  $t=0$, $r=0$ spin flip fractionalizes, it spreads out to neighbor sites, reaching them at a time defined by the spinon light cone velocity $v$. If one traces the light cones from each neighboring site in Fig.~\ref{fig:schematic}(a), one finds constructive interference on even neighbor sites: as each pair of spinons passes through a site, the spin flips (two spinons of the same kind produce a spin flip), and again at a later time $2a/v$, and so on, creating a sinusoidal oscillation in the spin correlators. Meanwhile, on the odd sites one finds destructive interference between spinons and antispinons at all times. From this perspective $Re[G(r,t)]=0$ on the odd neighbor sites is a direct observation of particle-antiparticle annihilation.

% Both these pictures are valid ways of understanding $G(r,t)$ and by extension, $S(Q,\omega)$. The latter interpretation---particle-antiparticle interference---gives a new meaning to the $Q=\pi$ spinon continuum as a direct result of spinon-antispinon annihilation on the odd sites.

%\paragraph{Changes with increasing temperature:}
%Finally, we turn to the evolution of $G(r,t)$ with temperature in Fig. \ref{fig:hightemp}. The first noticeable change gradual loss of oscillations in the quantum wake. This also can also be understood via spinon/antispinon light cones.
This spinon heuristic interpretation can explain the temperature evolution of $G(r,t)$ in KCuF$_3$.
As temperature increases, the static spin correlations and spin entanglement are suppressed~\cite{scheie2021entanglement}, which destroys the coherence of the spinons from neighboring sites as illustrated in Fig.~\ref{fig:schematic}(b), and the oscillations vanish. %This explains why the $r=0$ oscillations quickly die out at higher temperatures, because the further neighbor spins' quasiparticles are not coherent with those of $r=0$. The same thing occurs for the further neighbor sites, such that no time-like correlations (oscillatory or otherwise) exist beyond $r=0$ (Fig.~\ref{fig:hightemp}(h)).

This also explains the shift to a ferromagnetic wavefront at high temperatures [Fig.~\ref{fig:hightemp}(h)].
At low temperatures, the spinons propagate atop a substrate of antiferromagnetic correlations, giving rise to antiferromagnetic oscillating interference patterns.
At higher temperatures, the static correlations are mostly gone and so are coherence with neighboring sites (evidenced by the vanishing $Im[G(r,t)]$), so the propagating spinons simply appear as a pair of up-spins hopping through the lattice.
In this way, the high temperature quantum wake directly shows spinon quasiparticles---one can ``see'' them in the data.
It is striking that a diffuse high-temperature $S(Q,\omega)$ could yield such a clear quasiparticle signature in $G(r,t)$. This technique could have profound implications for identifying exotic quasiparticles in other magnetic systems.

\section*{Conclusions:}

In conclusion, we have shown using KCuF$_3$ scattering that it is possible to resolve real-time spin dynamics of a local quantum quench via neutron scattering. This reveals details about the quantum dynamics which were not obvious otherwise. First, we are able to directly observe the formation of an orthogonal state within the quantum wake as the light cone scrambles the initial state, leaving behind decaying period-doubled $\pi/2$ oscillations. Second, using the imaginary $G(r,t)$ we observe quantum coherence as revealed by non-commuting observables between spins more than 10 neighbors distant in Fig. \ref{fig:fouriertransform_imag}. This is far longer range ``quantumness'' than is revealed by entanglement witnesses \cite{scheie2021entanglement}. Third, the high-temperature $G(r,t)$ shows the spinon quasiparticles visually in the data, without need for theoretical models. Such details are  difficult or impossible to see with other techniques. %Neutron scattering holds a distinct advantage over cold-atom or trapped-ion experiments ~\cite{Langen2013,cheneau2012light,jurcevic2014quasiparticle} because it can probe equilibrium quantum dynamics in the thermodynamic limit, and also yield imaginary spin correlations. These both provide key information: the high temperature $G(r,t)$ directly shows spinon quasiparticles, and the imaginary component shows the time-evolution of mutual incompatibility. Thus in a very real sense, the spin-flip wavefront leaves Heisenberg uncertainty in its wake.

%We have also used this new perspective on spin-spin correlation to show key features of the 1D Heisenberg spin chain quantum wake, whereby stationary spinons produce quasiperiodic $\pi/2$ oscillations extending to long times.

The ability to probe short time and space dynamics of quasiparticles is of key importance to both fundamental quantum mechanics research and technological applications. On the fundamental side, the existence of a quantum wake with quasiperiodic $\pi/2$ oscillations shows behavior not captured by bosonization, which means theorists need to re-tool their analytic methods to understand the short-time dynamics of quantum spin chains. Also, measuring $G(r,t)$ at a well-defined finite temperature may shed light on eigenstate thermalization and quantum scrambling in higher-dimensional systems.
On the applications side, $G(r,t)$ is more closely related to the output of current quantum computers and so may provide more direct application of this technology. Also, understanding the short-time behavior of quasiparticles in quantum systems is a crucial step in using them for quantum logic operations in real technologies. Neutron scattering derived $G(r,t)$ provides key insight into these problems.

\section*{Methods}

Full methods are available in the Supplementary Information.

\subsection*{Extracting \texorpdfstring{$G(r,t)$}{G(r,t)} from inelastic neutron scattering}

The high-energy scattering data was measured on MAPS at ISIS with phonons subtracted, and low energy ($<7$~meV) scattering data at high temperatures---where the MAPS data is noisy---was filled in with data measured on SEQUOIA~\cite{Granroth2006} at ORNL's SNS~\cite{mason2006spallation}. Both data sets were corrected for the magnetic form factor, and the resulting combined data are shown in Fig.~\ref{fig:hightemp}.

We then masked the elastic scattering (as it is mostly nonmagnetic incoherent scattering), calculated the negative energy transfer scattering using detailed balance, and computed the Fourier transform of the neutron scattering data in both $Q$ and $\hbar \omega$, yielding spin-spin correlation in real space and time $G(r,t) = \langle S(0) \cdot S_r(t) \rangle$. (Prior to transforming, the high energy MAPS data was interpolated using Astropy Gaussian interpolation~\cite{Astropy2018} to create a uniform grid.)

The short-distance long-time $G(r,t)$ dynamics are governed by the lowest measured energies. In this case, the low energy cutoff was $0.7$~meV which means $G(r,t)$ is reliable only up to $\sim 5\times10^{-13}$~s.  Further details are given in the Supplemental Information. Thus, the long-time dynamics are inaccessible to the current data set.
This being said, there is an important visible difference between KCuF$_3$ and the Bethe Ansatz $G(r,t)$ at long times: KCuF$_3$ tends toward antiferromagnetic correlations (odd neighbors fade towards red, even neighbors fade more blue), whereas the Bethe ansatz shows no such trend. This is because KCuF$_3$ is magnetically ordered at 6~K due to interchain couplings, and thus has an infinite-time static magnetic pattern; but the idealized 1D Heisenberg AFM does not. Remarkably, the Van Hove function picks this up even though the elastic line---and thus the Bragg intensity---was not included in the transform.

\subsection*{Theoretical simulations}

The Bethe Ansatz plots were produced from data obtained using the ABACUS algorithm \cite{2009_Caux_JMP_50} which computes dynamical spin-spin correlation function of  integrable models through explicit summation of intermediate state contributions as computed from (algebraic) Bethe Ansatz.
Linear spin wave calculations were carried out using SpinW~\cite{SpinW}.

In the Supplemental Information, we also consider (i) the $S=1/2$ ferromagnet using both density matrix renormalization group theory (DMRG) and LSWT, and (ii) the quantum $S=1/2$ Ising spin chain for various anisotropies using perturbation theory. %~\cite{Ishimura_1980}.

%\section*{Data Availability}
%
%All data is available at [insert URL when available].

% \bibliography{RSCorrelations}


\subsection*{Acknowledgments}
We are thankful to Takeshi Egami for enlightening discussions. The research by P.L. was supported by the Scientific Discovery through Advanced Computing (SciDAC) program funded by the US Department of Energy, Office of Science, Advanced Scientific Computing Research and Basic Energy Sciences, Division of Materials Sciences and Engineering. This research used resources at the Spallation Neutron Source, a DOE Office of Science User Facility operated by the Oak Ridge National Laboratory. JSC acknowledges support from the European Research Council (ERC) under ERC Advanced grant 743032 DYNAMINT. The work by DAT and SEN is supported by the Quantum Science Center (QSC), a National Quantum Information Science Research Center of the U.S. Department of Energy (DOE).

\newpage

\quad 

\newpage

	\renewcommand{\thefigure}{S\arabic{figure}}
\renewcommand{\thetable}{S\arabic{table}}
\renewcommand{\theequation}{S.\arabic{equation}}
\renewcommand{\thepage}{S\arabic{page}}  
\setcounter{figure}{0}
\setcounter{page}{1}
\setcounter{equation}{0}

\section*{Supplemental Information for Quantum wake dynamics in Heisenberg antiferromagnetic chains}

\section{Real and imaginary \texorpdfstring{$G(r,t)$}{G(r,t)}}

% Imaginary $G(r,t)$ requires quantum behavior because the quantum limit allows for an imbalance in the positive and negative energy transfer scattering at finite temperatures, which produces nonzero imaginary values under 2D Fourier transform.  In the limit of $S \rightarrow \infty$ or $T \rightarrow \infty$, the positive and negative energy transfer are equivalent, yielding a completely real $G(r,t)$.

As noted in the main text, the real and imaginary parts of $G(r,t)$ probe different quantum mechanical functions.
The imaginary $G(r,t)$ is written
\begin{equation}
Im[G(r,t)] = \frac{1}{2i}\big[\langle S_i^z(0) S_j^z(t) \rangle - \langle S_i^z(0) S_j^z(t) \rangle^* \big]
\end{equation}
$$
= \frac{1}{2i}\langle S_i^z(0) S_j^z(t) - S_j^z(t) S_i^z(0) \rangle
$$
which can be written with a commutator
\begin{equation}
Im[G(r,t)] = \frac{1}{2i}\langle [S_i^z(0), S_j^z(t)] \rangle     \label{eq:imag}
\end{equation}
%which is NOT the equation for the retarded (dissipative) response function~\cite{Lederer2020}. 
Therefore, the imaginary component of $G(r,t)$ directly gives the dissipative susceptibility.
Following the same derivation, we arrive at the equation for the real part of $G(r,t)$ 
\begin{equation}
Re[G(r,t)] = \frac{1}{2}\langle \{S_i^z(0), S_j^z(t)\} \rangle.  \label{eq:real}
\end{equation}

Comparing eq.~\eqref{eq:imag} and eq.~\eqref{eq:real}, one can see why the imaginary part of $G(r,t)$ goes to zero at infinite temperature or in the classical limit: as all states are equally populated, the commutator (and thus dissipations) vanish. This corresponds to $\mathcal{S}(-q,-\omega) = \mathcal{S}(q,\omega)$.
Meanwhile, so long as correlations exist, eq.~\eqref{eq:real} is nonzero even at infinite temperature or in the classical limit.

A nonzero commutator between spins has a non-trivial relationship to quantum entanglement.
Generically, the equal time spin operators of any two different spins always commute: $[S_i^{\alpha}(0),S_j^{\beta}(0)] = 0$, no matter whether the wavefunction formed by the two spins has off-diagonal density matrix components (i.e., no matter whether the two spins are entangled). To obtain a nonzero commutator (and thus an uncertainty relation), one must introduce time evolution to one of the spins with a Hamiltonian that involves interaction between $S_i$ and $S_j$. In this case, the commutator may be nonzero. %If the Hamiltonian used to time-evolve the system does not involve any interaction between $S_i$ and $S_j$, the commutator is always zero. Thus time evolution and magnetic interaction are key ingredients to a nonzero commutator.

%The relationship between this out of time spin commutator and quantum entanglement between $S_i$ and $S_j$ is non-trivial. 
The presence of Heisenberg uncertainty generically implies quantum coherence between two operators, such that an observation of one quantity destroys the other's state. This is actually the opposite of quantum entanglement, where observation of one quantity determines the other's state. Thus, the presence of nonzero imaginary $G(r,t)$ does not necessarily imply quantum entanglement (defined by off-diagonal density matrix components), but instead it witnesses a quantum coherence between $S_i$ and $S_j$. This is related (but not formally equivalent to) quantum discord, which is a generic measure of quantum correlations \cite{Ollivier_2001,Luo_2008}. 
Thus $Im[G(r,t)]$ is a witness of the quantum coherence of a system, which in the case of KCuF$_3$ extends to beyond 10 neighbors along the chain at 6~K. This is in accord with its highly coherent and entangled ground state. As temperature increases, the imaginary $G(r,t)$ becomes severely truncated in space, as shown in the main text Fig. 1.

\section{Ferromagnetic spin chain}\label{app:FM}

\begin{figure}
	\centering
	\includegraphics[width=0.46\textwidth]{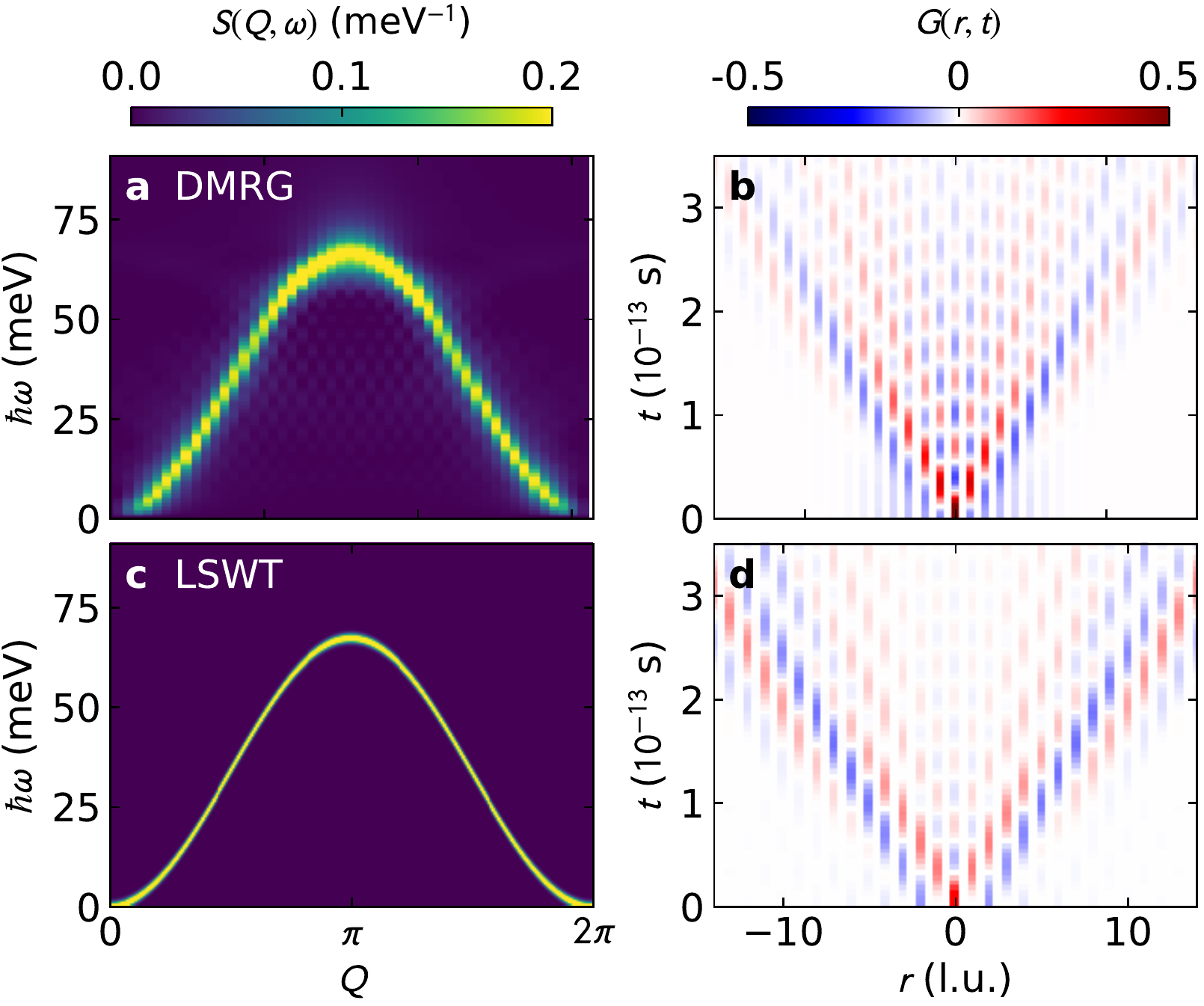}
	\caption{Real space spin correlations for a 1D Heisenberg ferromagnetic $S=1/2$ chain at $T=0$, simulated with DMRG and LSWT. Simulated neutron spectra are shown on the left, and Van Hove spin correlations (real part) are on the right. In this case, the semiclassical LSWT spin correlations are close to the DMRG quantum calculations, but the DMRG shows more oscillations near $r=0$ at long times.}
	\label{fig:ferromaget}
\end{figure}

As discussed in the main text, the $\pi/2$ stationary oscillations inside the quantum wake can be understood heuristically as spinon-antispinon interference. Here we propose an alternative (equally valid) heuristic for understanding the $\pi/2$ oscillations within the quantum wake: the effects of a spin-down operator on a down spin. If the $t=0,r=0$ spin is flipped up-to-down and the down-spin spinon propagates outward, the spin-lowering operator acting on a down spin results in zero. Meanwhile, the spin-lowering operator acting on an up-spin results in a spin flip. Thus odd (up-spin) sites correlations go to zero as the spinon light cone passes, and even sites flip.

To confirm the validity of these spinon heuristics, we also consider the isotropic $S=1/2$ ferromagnetic chain, and simulate its $T=0$ neutron spectra with DMRG \cite{PhysRevLett.69.2863, PhysRevB.48.10345, Alvarez2009} and LSWT, see Fig.~\ref{fig:ferromaget}. The DMRG calculation was performed on a chain of $L=50$ sites with open boundaries, keeping up to $m=500$ states in the calculation. $S(Q,\omega)$ was calculated using the DMRG++ \cite{Alvarez2009} implementation of the Krylov-space correction vector method \cite{PhysRevB.60.335, PhysRevE.94.053308}, and a Lorentzian energy broadening with half-width at half-maximum (HWHM) $\eta=0.1|J|$ to account for the finite-size system. To isolate the inelastic scattering, a Lorentzian with height $S(Q,0)$ was substracted at each $Q$-point.

Unlike the AFM case, excitations from the zero temperature FM ground state are spin flips of the same direction, which would mean no antiparticles are created and no destructive interference will occur. This is indeed what we see: all sites oscillate in time above the light cone, and no continuum exists in $S(Q,\omega)$.

If there were regular destructive interference, it would by necessity create a continuum in the neutron spectrum $S(Q,\omega)$: well-defined oscillations in time corresponds to a sharp mode in energy, whereas suppressed (or quickly decaying) correlations correspond to diffuse modes in energy. So even without transforming the neutron data into $S(Q,\omega)$, it should be obvious from the well-defined mode that there is no significant particle-antiparticle annihilation in $G(r,t)$ for the zero temperature FM spin chain.

\section{Toward the Ising limit}\label{app:Ising}

\begin{figure}
	\centering
	\includegraphics[width=0.47\textwidth]{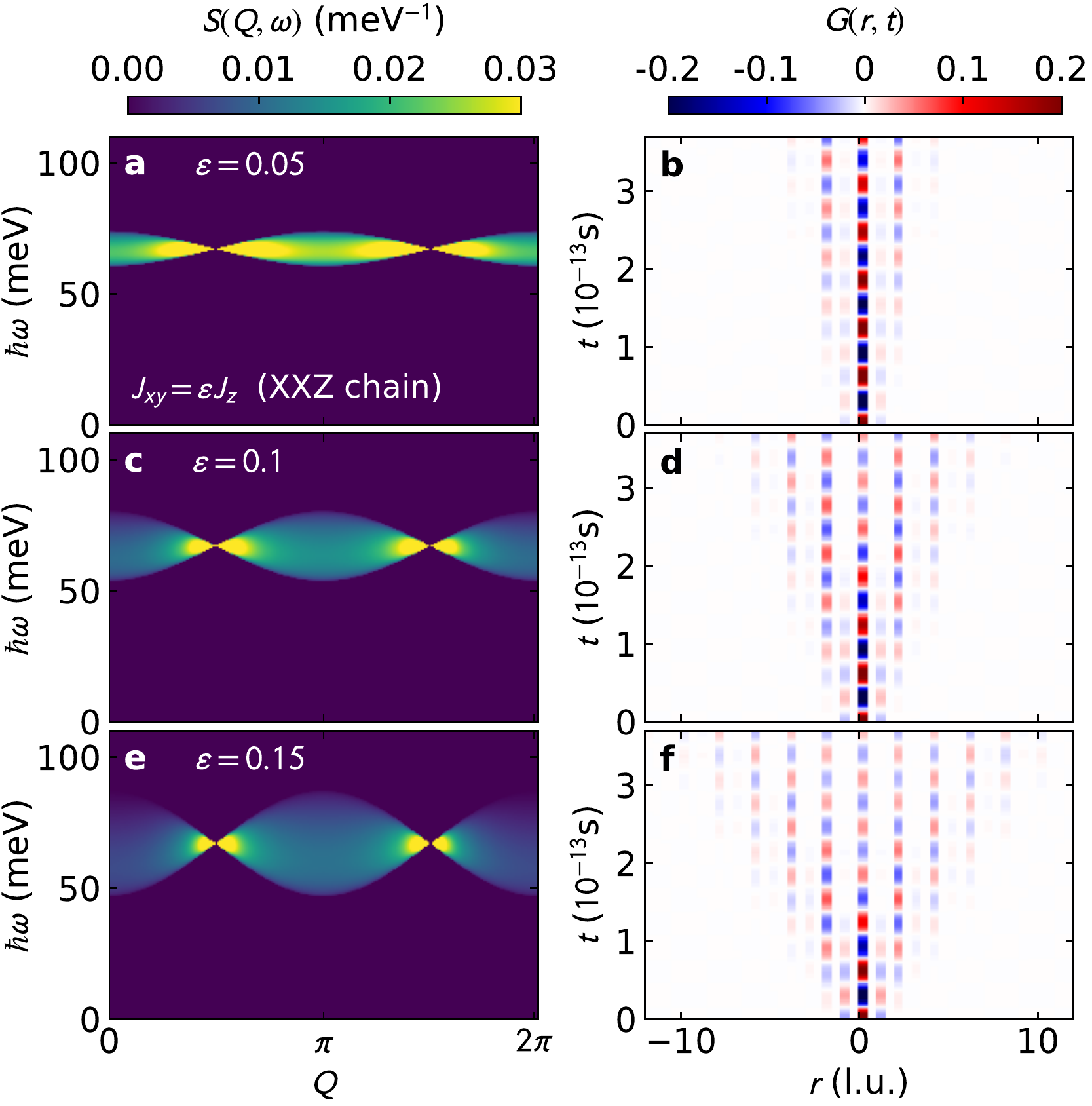}
	\caption{Simulated spin correlations for a 1D Ising AFM chain for three different values of anisotropy. Simulated $S_{xx}(Q,\omega)$ neutron spectra are shown on the left column (calculated via perturbation theory as described in Ref. \cite{Ishimura_1980}), and Van Hove spin correlations (real part only) are on the right column. Similar to Fig. 2 in the main text, the odd neighbor sites correlations decay to zero while even neighbor sites oscillate to long times. The ``light cone'' gets steeper and steeper as the Ising limit is approached.}
	\label{fig:Iising}
\end{figure}

Figure \ref{fig:Iising} shows the calculated real space correlations from perturbation theory at $T=0$ approaching the Ising limit. The $S(q,\omega)$ was calculated as described in Ref. \cite{Ishimura_1980} and transformed into $G(r,t)$. There are several things worth noting: first, just like the $S=1/2$ Heisenberg chain, the odd neighbor sites' correlations go to zero, in accord with spinon-antispinon interference.
Second, there is no well-defined wavefront visible in the data---possibly because the simulated intensity only includes the inelastic channel.
Finally, as the Ising limit is approached, the ``light cone'' gets steeper and steeper, corresponding to slower and slower spinon velocities.

\section{The XY limit}\label{app:XY}
\begin{figure*}
	\includegraphics[width=\textwidth]{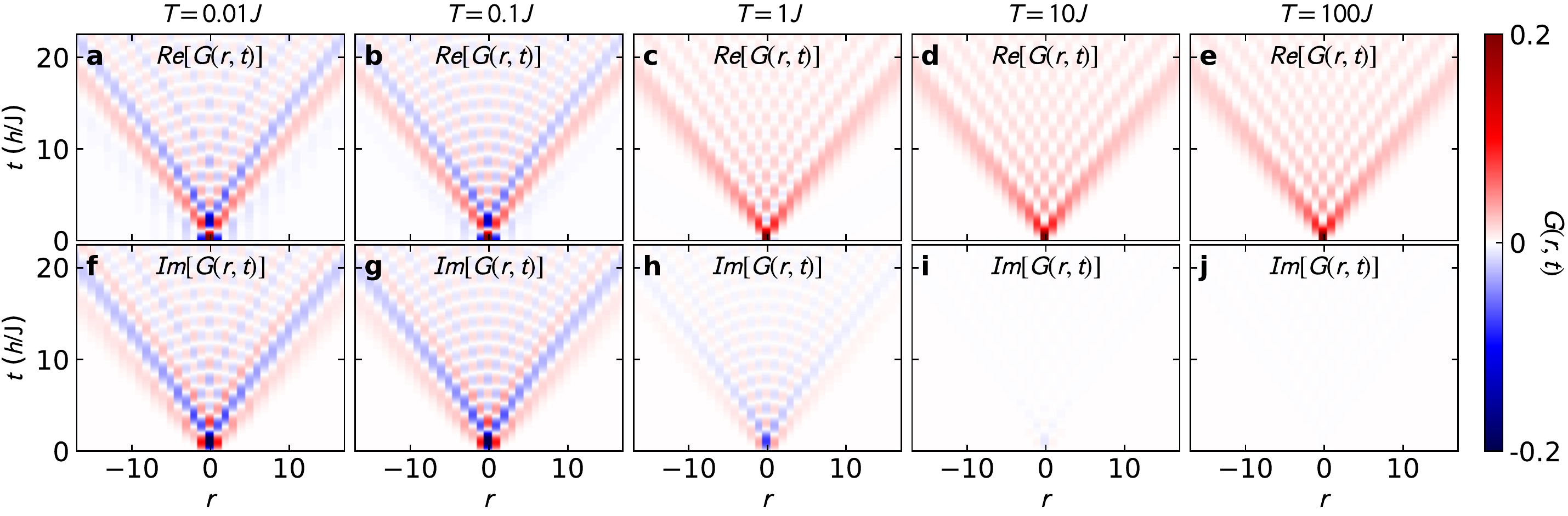}
	\caption{\label{fig:XX:SzSzCorr}Real (top row) and imaginary (bottom row) part of $\left\langle S_j^z (t) S_l^z(0) \right\rangle$ for the  XX model, Eq.~\eqref{eq:XX:longitudinal}, as a function of temperature. The real part becomes non-negative as $T$ increases, whereas the imaginary part vanishes. In addition, we see a lightcone with similarly oscillatory behavior as in the Heisenberg model.}
\end{figure*}
Although the isotropic Heisenberg chain model applicable to KCuF$_3$ can be solved exactly using the Bethe ansatz, the resulting expressions are often complicated. We can instead consider the antiferromagnetic isotropic XY-model (or XX-model) \cite{Lieb1961},
\begin{equation}
\mathcal{H}=J\sum_{i=1}^{N}\left[S_i^x S_{i+1}^x + S_i^y S_{i+1}^y\right],    \label{eq:XY}
\end{equation}
for which simpler, closed-form expressions can be obtained using the Jordan-Wigner formalism. At zero magnetic field, for a chain of $N$ sites with open boundary conditions, the longitudinal dynamical correlation between two lattice sites $j$ and $l$ can be written \cite{Goncalves1980,Cruz1981}
\begin{widetext}
	\begin{align}
	\left\langle S_j^z (t) S_l^z(0) \right\rangle	&=	\frac{1}{\left( N+1 \right)^2} \left[ \sum_k \left( \sin^2 \left( kj\right) \right) \tanh \left( \frac{J\cos k}{2k_B T}\right) \right] \times \left[ \sum_k \left( \sin^2 \left( kl\right) \right) \tanh \left( \frac{J\cos k}{2k_B T}\right) \right]	\nonumber\\
	&+ \frac{1}{\left( N+1 \right)^2} \left( \left[\sum_k \sin \left( kj\right)\sin \left( kl\right) \left\{ \cos \left(  tJ\cos k\right) -i\sin\left(tJ\cos k\right)\tanh \left( \frac{J\cos k}{2k_B T} \right)\right\} \right]^2 \right)	\nonumber\\
	&- \frac{1}{\left( N+1 \right)^2} \left( \left[\sum_k \sin \left( kj\right)\sin \left( kl\right) \left\{ i\sin\left(tJ\cos k\right) - \cos \left(  tJ\cos k\right) \tanh \left( \frac{J\cos k}{2k_B T} \right)\right\} \right]^2 \right),	\label{eq:XX:longitudinal}
	\end{align}
	where $k = \frac{m \pi}{N+1},	\quad	1\le m\le N+1$, is the momentum. Due to their simple structure, these sums can be evaluated at arbitrary times, temperatures and finite sizes. Yet they still capture several of the qualitative features observed in the KCuF$_3$ $G(r,t)$, as shown in Fig.~\ref{fig:XX:SzSzCorr}. It is easy to analytically see the emergence of real-valued ferromagnetic correlations for all times $t$ by considering the high-temperature limit of Eq.~\eqref{eq:XX:longitudinal}, where the $\tanh$ factors vanish, leaving
	\begin{align}
	\left\langle S_j^z (t) S_l^z(0) \right\rangle	&\approx \frac{1}{\left( N+1 \right)^2} \left( \left[\sum_k \sin \left( kj\right)\sin \left( kl\right) \left\{ \cos \left(  tJ\cos k\right) \right\} \right]^2 + \left[\sum_k \sin \left( kj\right)\sin \left( kl\right) \left\{ \sin\left(tJ\cos k\right) \right\} \right]^2 \right),
	\end{align}
	which is manifestly real and non-negative for all times $t$. In the thermodynamic limit we have the expressions \cite{Goncalves1980},
	\begin{align}
	\left\langle S_j^z (t) S_l^z(0) \right\rangle	&=	\left\{ \begin{array}{cc} \frac{1}{4} \left[ J_{j-l}\left( Jt\right) - \left( -1\right)^l J_{j+l}\left(Jt\right) \right]^2,&	\quad\, \text{for }T=\infty, \\ \frac{1}{4} \left[ F_{j-l}\left( Jt\right) - \left( -1\right)^l F_{j+l}\left(Jt\right) \right]^2,& \quad \text{for }T=0,\end{array}\right.	\label{eq:corrs:thermodynamiclimit}
	\end{align}
	where $J_n$ is the $n$th Bessel function of the first kind and $F_n=J_n+iE_n$, where $E_n$ is the Weber function.
\end{widetext}

\section{Time-limit of reliability}\label{app:timelimit}

\begin{figure}
	\centering
	\includegraphics[width=0.44\textwidth]{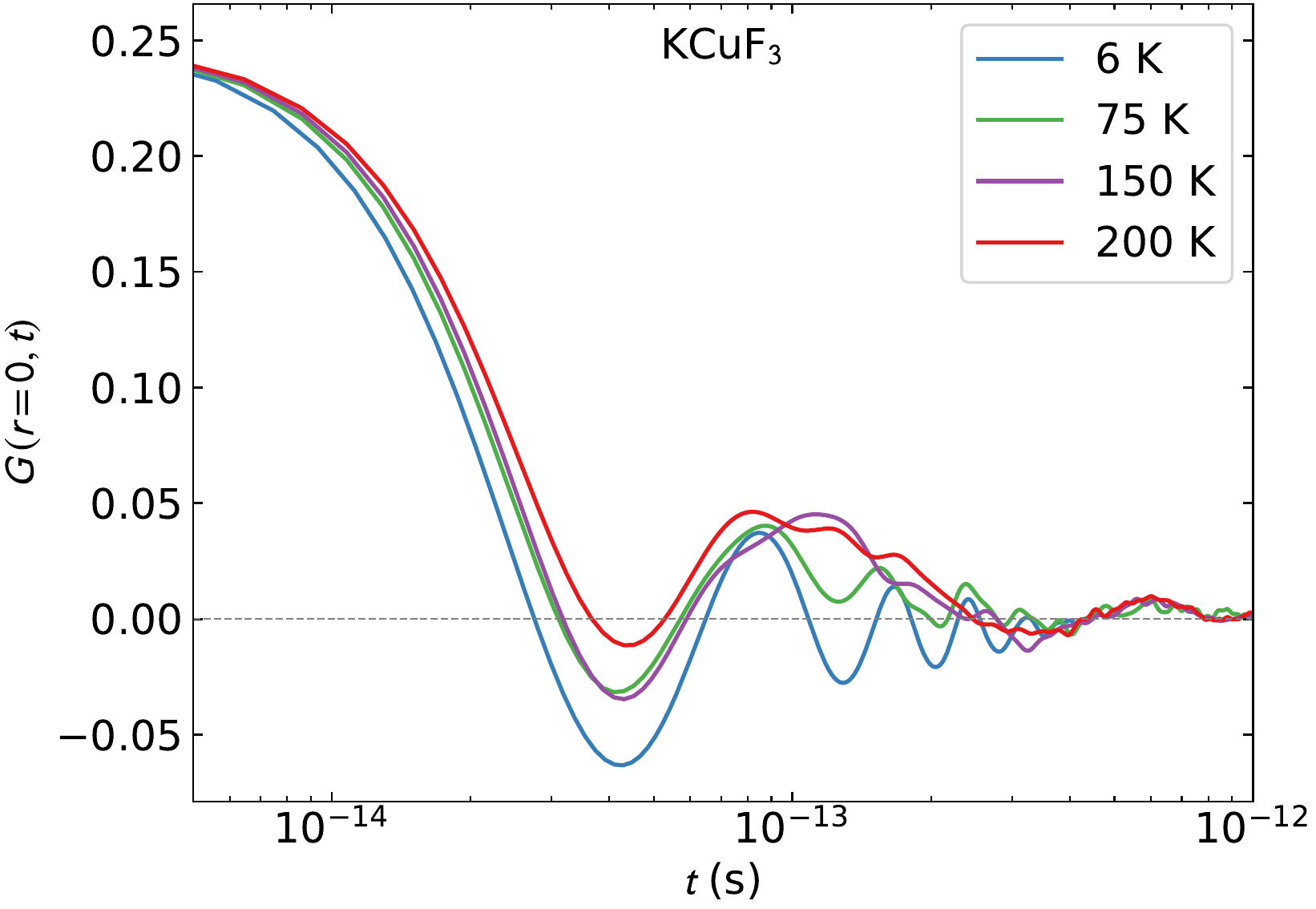}
	\caption{On-site correlation $r=0$ for KCuF$_3$ at various temperatures, showing the oscillations decaying. Beyond $4\times 10^{-13}$~s, the results are not reliable and the ``ringing'' from the low-energy cutoff begins to dominate the signal.}
	\label{fig:R0}
\end{figure}

The long time dynamics of a Fourier transform is determined by the lowest frequencies. Consequently, the reliability of the calculated $G(r,t)$ at long times is governed by the lowest measured energy. In this case, the low-energy cutoff from the SEQUOIA experiment was 0.7~meV, which yields a cutoff in time of $\frac{h}{\hbar \omega} = 6 \times 10^{-12}$~s, where $h$ is Planck's constant. However, the boundary between the MAPS and SEQUOIA data is 7 meV, which yields a slight artifact in the data and causes the $G(r,t)$ to ``ring'' with a period $\frac{h}{\hbar \omega} = 6 \times 10^{-13}$ s---this behavior is an artifact and is not physical. When the calculations are safely below this threshold, the $G(r,t)$ is reliable, as shown in Fig. \ref{fig:R0}. 
Although the lower energy SEQ data produced this ringing, we found that it was necessary to include in order to get a clean Fourier transform signal at higher temperatures.

To be completely safe from ringing effects, we find that one needs to stay below half the cutoff time ($3 \times 10^{-13}$~s for this experiment). Thus, any experiments aiming to measure $G(r,t)$ to long times must measure to appropriately high resolution and low energies.

\section{Quantum scrambling and out of time correlators}

Quantum scrambling is typically studied using out of time order correlators (OTOC)~\cite{maldacena2016bound,Swingle2018,Garttner2017,Li_2017,Luitz_2017,Colmenarez_2020}, which in spin chains are defined as
\begin{equation}
F(t)= \langle \hat{S}_a^{\dagger}(t) \hat{S}_b^{\dagger} \hat{S}_a(t) \hat{S}_b \rangle
\end{equation}
where $\hat{S}_a(t)$ and $\hat{S}_b$ are two different spin operators at time $t$ and $t=0$ respectively. The OTOC is related to the commutator between these operators $Re [F(t)] = 1 - \langle |[\hat{S}_a(t), \hat{S}_b]|^2 \rangle /2$, which functionally makes the OTOC a measure of how $\hat{S}_a(t)$ and $\hat{S}_b$ fail to commute~\cite{Garttner2017}. In 1D spin chains, OTOCs reveal quantum scrambling above the light cone~\cite{Nakamura_2019,Luitz_2017,Kim_2013}.
This is similar (but not identical) to imaginary $G(r,t)$, which also measures  $[\hat{S}_a(t), \hat{S}_b]$ (Eq.~\ref{eq:imag}), and thus provides similar information. %From this perspective, the propagating spinon wavefront scrambles the spin chain system as it passes each site.

As shown in Fig.~\ref{fig:hightemp_imag} and Fig. 2 of the main text, imaginary $G(r,t)$ correlations are only nonzero above the light cone, in good agreement to the commuting spin operators in the Heisenberg antiferromagnetic ground state. Above the light cone, both Bethe ansatz and KCuF$_3$ show nonzero negative static $\mathrm{Im}[G(r,t)]$ on the odd sites, and oscillating but average positive $\mathrm{Im}[G(r,t)]$ on the even sites. This concurs with quantum scrambling, where time-like separated spin operators do not commute with the original magnetism.

% On a more conceptual level, the $\mathrm{Im}[G(r,t)]$ shows how the system settles into an orthogonal state: the positive and negative average values on the even and odd sites respectively show that the spin quantization axis has rotated from the original axis and has settled into an antiferromagnetic correlated state at $\pi/2$ rotation from the original $z$ axis. Thus, it appears that the imaginary $G(r,t)$ not only reveals the dissipations and uncertainty relations in real space, but also correlations orthogonal to the $t=0$ state.

% \bibliography{RSCorrelations}


\end{document}